\documentstyle[amssymb,aps,epsf,12pt]{revtex}

\begin{document}
\author{Alexander Vilenkin and Serge Winitzki}
\title{Probability distribution for $\Omega $ in open-universe inflation}
\date{\today }
\maketitle

\begin{abstract}
The problem of making predictions in eternally inflating universe that
thermalizes by bubble nucleation is considered. A recently introduced
regularization procedure is applied to find the probability distribution for
the ensemble of thermalized bubbles. The resulting probabilities are shown
to be independent of the choice of the time parametrization. This formalism
is applied to models of open ``hybrid'' inflation with $\Omega <1$.
Depending on the parameters of the model, the probability distribution for $%
\Omega $ is found to have a peak either very close to $\Omega =1$, or at an
intermediate value of $\Omega $ in the range $0.03\lesssim \Omega <1$.
\end{abstract}

\section{Introduction}

A flat universe with $\Omega =1$ is usually regarded as a firm prediction of
inflationary models \cite{Flatness}. However, observational evidence for a
flat universe is far from being certain, and progressively more precise
estimates of the matter density of the Universe may yet show that the value
of $\Omega $ is less than $1$. Not surprisingly, theorists found ways of
modifying the models to make them compatible with $\Omega <1$ \cite
{OpenInfl,Turok1,OpenInfl1,LindeO1,Density}. In the new class of models,
called ``open universe inflation'', the inflaton potential has a metastable
minimum separated from the true vacuum by a potential barrier. The false
vacuum decays through bubble nucleation, and the inflaton field rolls
towards the true vacuum inside the bubbles, while inflation continues
outside. Co-moving observers inside a bubble would, after thermalization of
the inflaton, see themselves in an open homogeneous universe with $\Omega <1$%
. In the usual inflationary models, flatness ($\Omega \approx 1$) results
from a large amount of inflation needed for the observed homogeneity of the
universe. However, in models based on bubble nucleation, the homogeneity of
the open universe inside a bubble is ensured by the symmetry of the bubble.
So, the number of $e$-foldings of inflation after nucleation (typically, of
order $60$) can be fine-tuned to give a specific value of $\Omega $ between $%
0$ and $1$.

Although the nucleating bubbles expand at speeds approaching the speed of
light, the false vacuum regions that separate them expand even faster. As a
result, inflation never ends and bubble nucleation continues {\em ad
infinitum}. If all nucleated bubbles are identical, then they will evolve to
thermalized regions with the same value of $\Omega $. (To compare the values
of $\Omega $ in different regions, we can evaluate them at a fixed reference
temperature, say, $T=2.7$K.) However, there may be several types of bubbles
giving rise to different values of $\Omega $, and in some models, like the
model of hybrid inflation considered by Linde and Mezhlumian \cite{LindeO1}, 
$\Omega $ can be a continuous variable. A natural problem in this kind of
model would be to find the probability distribution for $\Omega $. This
problem is the focus of the present paper.

Our approach will be based on the assumption that we are ``typical'' among
the civilizations inhabiting the universe. Here, the ``universe'' is
understood as the entire spacetime; our civilization is assumed to be
typical among all civilizations, including those that no longer exist and
those that will appear in the future. The assumption of being typical was
called the ``principle of mediocrity'' in Ref.\ \cite{Predictions}. It is a
version of the ``anthropic principle'' which has been extensively discussed
in the literature \cite{Anthropic,Anthropic1,Anthropic11,LB,Anthropic2}. In
this approach, the probability for us to observe a certain value of $\Omega $
is proportional to the total number of civilizations that will observe it.
The total number of civilizations in a co-moving region can be expressed as
the volume of that region ${\cal V}_{*}$ at thermalization multiplied by the
number $\nu _{\text{civ}}$ of civilizations that evolve per unit of
thermalized volume \cite{Fn3}. The ratio of probabilities for a ``typical''
observer to find oneself in regions of type $1$ and type $2$ is then given
by \cite{Predictions}: 
\begin{equation}
\frac{{\cal P}^{\left( 1\right) }}{{\cal P}^{(2)}}=\frac{{\cal V}%
_{*}^{\left( 1\right) }}{{\cal V}_{*}^{\left( 2\right) }}\frac{\nu _{\text{%
civ}}^{\left( 1\right) }}{\nu _{\text{civ}}^{\left( 2\right) }}.
\label{CivRatio}
\end{equation}

At this point the reader may be inclined to put this paper aside. What hope
can we have to estimate the number of civilizations if we do not understand
the conditions necessary for the evolution of life, let along consciousness?
However, we believe that the situation is not as bad as it may seem. In
models with a continuous spectrum of $\Omega $, like the model of Linde and
Mezhlumian, the nucleated bubbles have identical particle physics. The
difference in $\nu _{\text{civ}}\left( \Omega \right) $ is then due only to
the difference in the evolution of density fluctuations in bubbles with
different values of $\Omega $. Roughly speaking, $\nu _{\text{civ}}\left(
\Omega \right) $ is proportional to the density of galaxies formed in
bubbles with the corresponding value of $\Omega $, and its calculation does
not require any input from biochemistry. We shall attempt to estimate $\nu _{%
\text{civ}}\left( \Omega \right) $ in Sec.\ \ref{CIV}, but until then our
main goal is to develop a method for evaluating the volume ratios ${\cal V}%
_{*}^{\left( 1\right) }/{\cal V}_{*}^{\left( 2\right) }$.

\begin{figure}[tbh]
\epsfysize 2.5 cm \epsffile{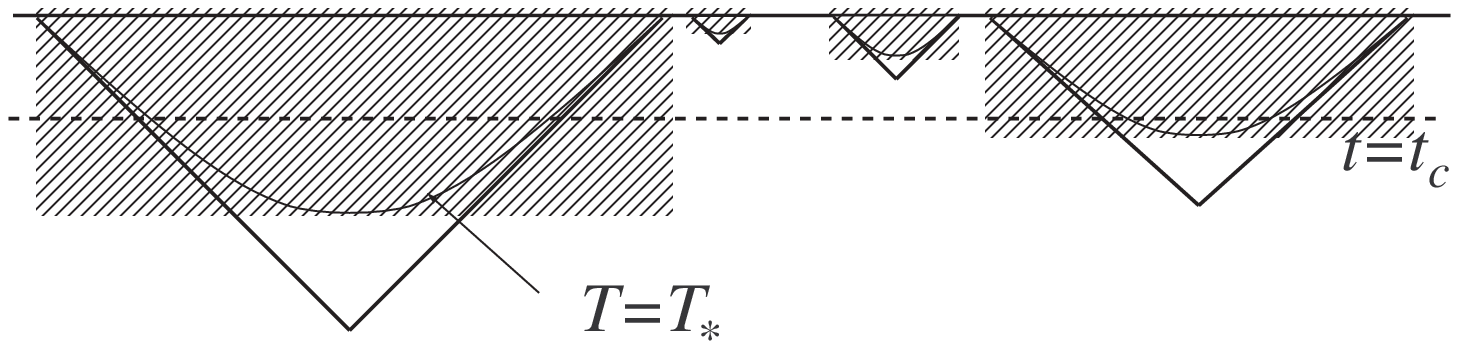}
\caption{A conformal diagram of bubbles nucleating in an inflating
background. The shaded regions of spacetime inside the bubbles are
thermalized. The thermalization surfaces are the boundaries of these
regions. They have an infinite $3$-volume which can be regularized by
introducing a cutoff hypersurface $t=t_c$ and keeping only the part of the
volume below this hypersurface.}
\label{BubFig0}
\end{figure}

The volume ${\cal V}_{*}^{(j)}$ is the combined $3$-volume of the
hypersurfaces of constant temperature $T=T_{*}$ inside bubbles of type $j$
(see Fig.\ \ref{BubFig0}). These thermalization hypersurfaces are spaces of
(approximately) constant negative curvature, and thus have infinite volume.
In order to define the volume ratio in Eq.\ (\ref{CivRatio}), this infinity
has to be regularized.

The most straightforward approach to regularization is to include in ${\cal V%
}_{*}$ only the part of the volume that thermalized prior to some time, $%
t=t_c$, and take the limit of the volume ratios ${\cal V}_{*}^{\left(
1\right) }/{\cal V}_{*}^{\left( 2\right) }$ as $t_c\rightarrow \infty $.
However, the result of this procedure is highly sensitive to the choice of
time variable $t$ (that is, to the choice of the cutoff surface) \cite
{LLM,GBL}. An alternative regularization prescription \cite
{MakingPredictions} is to cut off the volumes ${\cal V}_{*}^{(j)}$ at the
times $t_\epsilon ^{(j)}$ at which a small fraction $\epsilon $ of the
corresponding co-moving volumes is still in the inflating region. The
probability ratios (\ref{CivRatio}) are then defined by taking the limit of $%
{\cal V}_{*}^{\left( i\right) }/{\cal V}_{*}^{\left( j\right) }$ as $%
\epsilon \rightarrow 0$. The application of this procedure (which we shall
call the $\epsilon $-prescription) to models of stochastic inflation was
discussed in Refs. \cite{MakingPredictions,Uncertainties}, where it was
shown that the resulting probabilities are essentially independent of time
parametrization \cite{Fn1}.

In models of stochastic inflation, the inflaton field undergoes quantum
fluctuations on the horizon scale, and its evolution is described by a
diffusion equation \cite{Diffusion}. The physics of such models is very
different from that of bubble nucleation and expansion, and the methods of 
\cite{MakingPredictions,Uncertainties} are not directly applicable to
open-universe inflation. The purpose of this paper is to extend the results
of \cite{MakingPredictions,Uncertainties} to this case and, as an
application, to find the probability distribution for $\Omega $ in hybrid
inflation models.

The outline of the paper is as follows. In Sec.\ \ref{GEOM} we develop the
geometric formalism necessary to describe the thermalization hypersurfaces
within expanding bubbles. In Sec.\ \ref{PROB} we calculate the regularized
volume ratios in models with a discrete set of bubble types. Then in Sec.\ 
\ref{TIMES} we verify that the volume ratios thus obtained are independent
of time parametrization. In Sec.\ \ref{OMEGA} we extend the analysis to the
hybrid inflation model of \cite{LindeO1} with a continuous family of
bubbles. In Sec.\ \ref{CIV} we estimate the ``human factor'' $\nu _{\text{civ%
}}$ and calculate the probability distribution for $\Omega $. Conclusions
follow in Sec.\ \ref{CONCL}. Some calculations for Sections \ref{GEOM} and 
\ref{PROB} are presented in Appendices.

\section{Bubble geometry\label{GEOM}}

The goal of this Section is to find the $3$-volume of a thermalization
hypersurface cut off at a given time $t_\epsilon $. For simplicity, we shall
use proper time for calculations; it will be shown in Sec.\ \ref{TIMES} that
the resulting probabilities do not depend on the choice of the time variable.

In models of open-universe inflation, the inflaton potential $V\left(
\varphi \right) $ has a local minimum, $V=V_0$, corresponding to a
metastable false vacuum. In regions occupied by the false vacuum, the metric
is approximately de Sitter, 
\begin{equation}
ds^2=-dt^2+\exp \left( 2H_0t\right) \left[ dr^2+r^2d\Omega ^2\right] ,
\label{deSitterMetric}
\end{equation}
where $d\Omega ^2\equiv d\theta ^2+\sin ^2\theta d\phi ^2$ is the usual
spherical surface element, $H_0$ is determined by the false vacuum energy, $%
H_0=\sqrt{8\pi V_0/3}$, and we use the Planck units, $\hbar =c=G=1$.

\begin{figure}[tbh]
\epsfysize 5 cm \epsffile{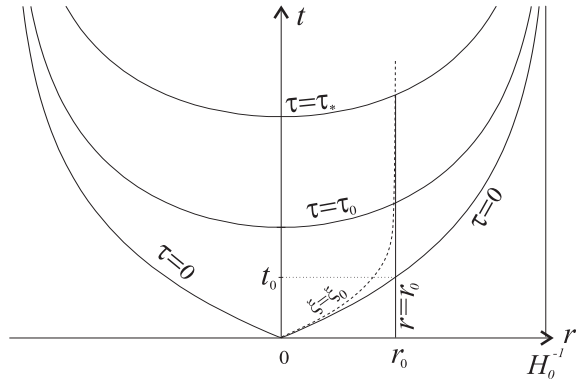}
\caption{Geometry of the bubble interior.}
\label{BubFig1}
\end{figure}

At the moment of nucleation, a spherical bubble is formed, with the inflaton
field in its interior on the other side of the potential barrier (with
respect to the false vacuum). The bubble then expands and the inflaton field
inside it evolves toward the true vacuum value, where it thermalizes. The
interior of the nucleated bubble looks like an open Robertson-Walker (RW)
universe in suitable coordinates $\left( \tau ,\xi \right) $, 
\begin{equation}
ds^2=-d\tau ^2+a^2\left( \tau \right) \left[ d\xi ^2+\sinh ^2\xi d\Omega
^2\right] .  \label{RWMetric}
\end{equation}
The scale factor $a\left( \tau \right) $ can be found from Einstein and
scalar field equations, which in the slow roll approximation take the form 
\begin{mathletters}
\begin{eqnarray}
\left( \frac 1a\frac{da}{d\tau }\right) ^2-\frac 1{a^2} &=&H^2\left( \varphi
\right) \equiv \frac{8\pi G}3V\left( \varphi \right) ,  \label{CosmEqu-a} \\
\frac{d\varphi }{d\tau } &=&-\frac 1{4\pi }H^{\prime }\left( \varphi \right)
.  \label{CosmEqu-phi}
\end{eqnarray}
Equations (\ref{CosmEqu-a})-(\ref{CosmEqu-phi}) are valid provided that 
\end{mathletters}
\begin{equation}
\left| \frac{H^{\prime }}{2\pi H}\right| \ll 1  \label{SlowRoll}
\end{equation}
and 
\begin{equation}
\left| H^{\prime }\right| \gg H^2.  \label{SmallFluct}
\end{equation}
Eq.\ (\ref{SlowRoll}) is the condition of slow roll, and Eq.\ (\ref
{SmallFluct}) ensures that quantum fluctuations are small, so that the
evolution of $a$ and $\varphi $ is essentially deterministic. The
coordinates $\tau $ and $t$ can be chosen so that the center of the
space-time symmetry of the bubble corresponds to $t=\tau =0$. Then the
surface $\tau =0$ is the future light cone of that center (see Fig.\ \ref
{BubFig1}). We assume that the initial bubble size is small on the horizon
scale $H_0^{-1}$, so that for our purposes the boundary of the bubble can be
approximated by this light cone.

The relation between the coordinates $\left( t,r\right) $ and $\left( \tau
,\xi \right) $ can be easily found if we assume, following \cite{Turok1},
that (i) the potential $V\left( \varphi \right) $ has nearly the same value
on the two sides of the barrier, and (ii) that the gravitational effect of
the bubble wall is negligible. (A similar, although more cumbersome,
calculation can be done for the more general case of non-negligible bubble
wall gravity and different expansion rates $H_0$, $H_1$ at the two sides of
the wall; see below and Appendix B for details.) Then, at sufficiently small
values of $\tau ,$ the geometry inside the bubble is close to that of de
Sitter space with the expansion rate $H_0$. The solution of Eq.\ (\ref
{CosmEqu-a}) with $H\left( \varphi \left( \tau \right) \right) \approx H_0$
is 
\begin{equation}
a\left( \tau \right) =\frac 1{H_0}\sinh \left( H_0\tau \right) .
\end{equation}
This is accurate as long as 
\begin{equation}
\left| \Delta H\right| \tau \approx \left| \frac{dH}{d\varphi }\frac{%
d\varphi }{d\tau }\tau \right| \tau =\frac{\left( H^{\prime }\left( \varphi
_0\right) \right) ^2}{4\pi }\tau ^2\ll 1,
\end{equation}
which gives 
\begin{equation}
\tau \ll \left| H^{\prime }\left( \varphi _0\right) \right| ^{-1}.
\label{TauSmallerThan}
\end{equation}
Here, $\varphi _0$ is the value of the field immediately after tunneling. At
times $\tau $ satisfying (\ref{TauSmallerThan}) the coordinates $\left( \tau
,\xi \right) $ are related to $\left( t,r\right) $ by the usual
transformation between spatially flat and open de Sitter coordinates: 
\begin{mathletters}
\label{NewCoords}
\begin{eqnarray}
t\left( \tau ,\xi \right) &=&\frac 1{H_0}\ln \left( \cosh H_0\tau +\sinh
H_0\tau \cosh \xi \right) ,  \label{NewCoords-t} \\
r\left( \tau ,\xi \right) &=&\frac 1{H_0}\frac{\sinh H_0\tau \sinh \xi }{%
\cosh H_0\tau +\sinh H_0\tau \cosh \xi }.  \label{NewCoords-r}
\end{eqnarray}
For $\tau \gtrsim \left| H^{\prime }\left( \varphi _0\right) \right| ^{-1}$,
Eqs.\ (\ref{NewCoords}) no longer apply, but the coordinates $\left(
t,r\right) $ can be continued to the entire bubble interior as co-moving
coordinates along the geodesics $r=const$.

Thermalization occurs at a hypersurface of equal RW time, $\tau =\tau _{*}$.
The time $\tau _{*}$ of thermalization is found from the evolution equation (%
\ref{CosmEqu-phi}): 
\end{mathletters}
\begin{equation}
\tau _{*}\approx \int_{\varphi _0}^{\varphi _{*}}\frac{d\tau }{d\varphi }%
d\varphi =-4\pi \int_{\varphi _0}^{\varphi _{*}}\frac{d\varphi }{H^{\prime
}\left( \varphi \right) },  \label{TauStar}
\end{equation}
where $\varphi _{*}$ is the value corresponding to the end of the slow roll
regime near the true vacuum. We shall assume that $H_0\tau _{*}\gg 1$.

The cutoff of the thermalization hypersurface at a time $t=t_\epsilon $
corresponds, in terms of the RW coordinates $\left( \tau ,\xi \right) $, to
cutting off the surface $\tau =\tau _{*}$ at some value $\xi =\xi _{*}$,
where $\xi _{*}$ is found from the requirement that the proper time $t$ at $%
\left( \tau _{*},\xi _{*}\right) $ be equal to $t_\epsilon $. Therefore, we
need to find the proper time along a geodesic $r=const$ which starts in the
false vacuum, continues into the bubble, and ends at $\left( \tau _{*},\xi
_{*}\right) $. This task is facilitated by the observation that the time $t$
along a co-moving geodesic in the de Sitter space after crossing the bubble
boundary (for $\exp \left( H_0\tau \right) \gg 1)$ becomes almost identical
to the RW time $\tau $ inside the bubble: 
\begin{equation}
t\left( \tau ,\xi \right) =\tau +\frac 2{H_0}\ln \cosh \frac \xi 2+O\left(
e^{-H_0\tau }\right) .  \label{TTau}
\end{equation}
In Appendix A it is shown that a co-moving geodesic $r=r_0$, after crossing
the bubble, rapidly approaches the RW co-moving geodesic line $\xi \left(
\tau \right) =const$. This, as well as Eq.\ (\ref{TTau}), holds for times $%
\tau $ within the range (\ref{TauSmallerThan}), when deviations of the
bubble interior from de Sitter space are small. But since the geodesics $%
r=const$ and $\xi =const$ nearly coincide at $\tau \gg H_0^{-1}$, it is
easily understood that Eq.\ (\ref{TTau}) is valid throughout the bubble
interior. Hence, the condition for the cutoff $\xi _{*}$ becomes:

\begin{equation}
\tau _{*}+\frac 2{H_0}\ln \cosh \frac{\xi _{*}}2=t_\epsilon .
\label{EquForXi}
\end{equation}
The solution of (\ref{EquForXi}) can be written as 
\begin{equation}
\xi _{*}=2\cosh ^{-1}\exp \frac{H_0\left( t_\epsilon -\tau _{*}\right) }2.
\label{XiStarSimple}
\end{equation}
The part of the thermalization hypersurface $\tau =\tau _{*}$ we are
interested in is bounded by $0\leq \xi \leq \xi _{*}$. Its $3$-volume,
calculated using the metric (\ref{RWMetric}), is 
\begin{equation}
V_{*}\left( t_\epsilon \right) =4\pi \int_0^{\xi _{*}}a^3\left( \tau
_{*}\right) \sinh ^2\xi d\xi =\pi a^3\left( \tau _{*}\right) \left( \sinh
2\xi _{*}-2\xi _{*}\right) .  \label{ThermVol}
\end{equation}

The calculations of the time cutoff (\ref{EquForXi}) were performed for the
case of unchanged expansion rate $H_1=H_0$ in the bubble interior
immediately after nucleation. The analogous cutoff condition for $H_1\neq
H_0 $ is derived in Appendix B. The thermalized volume $V_{*}\left(
t_\epsilon \right) $ as a function of $\xi _{*}$ is still given by (\ref
{ThermVol}).

\section{Regularized volume ratios\label{PROB}}

In this Section, we consider the situation where the bubbles come in several
types (labeled by $1$, $2$, etc.). We assume that the nucleation rates for
bubbles of each type are $\gamma _1$, $\gamma _2$, etc. A straightforward
generalization to a continuous variety of bubbles will follow in Sec.\ \ref
{OMEGA}. Our purpose is to find the thermalized volume ratios in bubbles of
different types. For that, we need to find the cutoff times $t_\epsilon
^{\left( 1,2\right) }$ and evaluate the ratio of volumes of thermalization
hypersurfaces regularized by cutoffs at $t_\epsilon ^{\left( 1,2\right) }$.
To simplify our calculations, we shall first consider nucleation of bubbles
of one type with nucleation rate $\gamma $, and subsequently generalize to
multiple types.

For a bubble that nucleates at time $t=0$, the regularized volume of
thermalization hypersurface is given by Eq.\ (\ref{ThermVol}) of the
previous section. Now we have to account for bubbles nucleated at all times,
starting for convenience at $t=0$. Bubbles will nucleate in spacetime
regions that are not already inside bubbles. (We disregard the possibility
of tunneling from the true vacuum back to the false vacuum.) A point $\left(
t_0,r\right) $ will {\em not} be inside a bubble if no bubbles were formed
in its past lightcone. The volume of the past lightcone of the point ($%
t_0,r=0$) in de Sitter spacetime is 
\begin{equation}
V_{lc}\left( t_0\right) =\int_0^{t_0}\frac{4\pi }3r_{lc}^3\left( t\right)
\exp \left( 3H_0t\right) dt,
\end{equation}
where $r_{lc}\left( t\right) $ is the null geodesic ending at time $t_0$ at $%
r=0$, 
\begin{equation}
r_{lc}\left( t\right) =\frac{\exp \left( -H_0t\right) -\exp \left(
-H_0t_0\right) }{H_0}.
\end{equation}
This gives 
\begin{equation}
V_{lc}\left( t_0\right) =\frac{4\pi t_0}{3H_0^3}-\frac{22}{9H_0^4}+O\left(
\exp \left( -H_0t_0\right) \right) .
\end{equation}
Therefore, for sufficiently late times $t\gg H_0^{-1}$, the probability for
a point $\left( t,r\right) $ not to be inside a bubble is 
\begin{equation}
P_{\text{outside}}\left( t\right) =\exp \left( -\gamma V_{lc}\right) =\exp
\left( -\frac{4\pi \gamma }{3H_0^3}t\right) ,  \label{ProbOutside}
\end{equation}
where we assumed that the nucleation rate is small, 
\begin{equation}
\gamma /H_0^4\ll 1,  \label{SmallGamma}
\end{equation}
and accordingly disregarded the factor $\exp \left( 22\gamma /9H_0^4\right) $%
.

The cutoff time $t_\epsilon $ is found from the condition that a fraction $%
\epsilon $ of co-moving volume is still inflating at that time. Since
inflation continues for some time inside the bubbles, the probability $%
P_{\inf }\left( t\right) $ of a point $\left( t,r\right) $ to be in a still
inflating region is not the same as the probability (\ref{ProbOutside}) of
being outside bubbles. If we assume that inflation inside bubbles lasts for
a period of proper time approximately equal to $\tau _{*}$ (the
thermalization time given by (\ref{TauStar})), then the points that are
still inflating at time $t$ are those which were outside bubbles at time $%
t-\tau _{*}$: 
\begin{equation}
P_{\inf }\left( t\right) \approx P_{\text{outside}}\left( t-\tau _{*}\right)
.  \label{ProbInf}
\end{equation}
Eq.\ (\ref{ProbInf}) is not exact because the proper time $t$ is different
from the time $\tau $ measured by the co-moving clocks inside the bubble;
however, this difference is not large because the co-moving geodesics that
define $t$ quickly approach the RW geodesics inside the bubbles soon after
they cross the boundaries. We will show in Appendix A that Eq.\ (\ref
{ProbInf}) is accurate as long as the nucleation rate is small as assumed in
(\ref{SmallGamma}).

Hence, the cutoff condition becomes 
\begin{equation}
P_{\inf }\left( t\right) =P_{\text{outside}}\left( t_\epsilon -\tau
_{*}\right) =\epsilon .  \label{CutoffCond}
\end{equation}
In (\ref{CutoffCond}), we can use the asymptotic formula (\ref{ProbOutside})
for $P_{\text{outside}}\left( t\right) $ because we will be taking the limit
of $\epsilon \rightarrow \infty $ for which $H_0\left( t_\epsilon -\tau
_{*}\right) \gg 1$.

Consider now a co-moving spatial volume equal to $CH_0^{-3}$ at $t=0$, where 
$C$ is a normalization constant corresponding to the initial number of
horizon-size regions. The total volume of regions outside bubbles at a later
time $t\gg H_0^{-1}$ is given by 
\begin{equation}
V_{\text{outside}}\left( t\right) =CH_0^{-3}\exp \left( 3H_0t\right) P_{%
\text{outside}}\left( t\right) \approx CH_0^{-3}\exp \left( dH_0t\right) ,
\label{VInf}
\end{equation}
where $d$ is the fractal dimension of the inflationary domain, 
\begin{equation}
d=3-\frac{4\pi \gamma }{3H_0^4}.
\end{equation}
We will later use the fact that $d\approx 3$.

The number of bubbles nucleated within the time interval $\left(
t_1,t_1+dt_1\right) $ is 
\begin{equation}
dN\left( t_1\right) =\gamma V_{\text{outside}}\left( t_1\right) dt_1,
\label{NBubbles}
\end{equation}
and therefore the combined thermalized volume inside all bubbles (from $t=0$
until the cutoff time $t_\epsilon $) is 
\begin{equation}
{\cal V}_{*}=\int_0^{t_\epsilon -\tau _{*}}V_{*}\left( t_\epsilon
-t_1\right) dN\left( t_1\right) =\int_0^{t_\epsilon -\tau _{*}}V_{*}\left(
t_\epsilon -t_1\right) \gamma V_{\text{outside}}\left( t_1\right) dt_1,
\label{ThermVolReg}
\end{equation}
where $V_{*}\left( t\right) $ is given by (\ref{ThermVol}). The integration
in (\ref{ThermVolReg}) is until $t_\epsilon -\tau _{*}$ because bubbles
nucleated after that time will not thermalize before $t_\epsilon $. We can
use the asymptotic formula (\ref{VInf}) for $V_{\text{outside}}\left(
t\right) $ since the integral in (\ref{ThermVolReg}) is exponentially
dominated by bubbles nucleated at late times.

Substituting (\ref{XiStarSimple}), (\ref{ThermVol}) and (\ref{VInf}) into (%
\ref{ThermVolReg}), we obtain: 
\begin{eqnarray}
{\cal V}_{*} &=&C\frac{\pi \gamma }{H_0^3}a^3\left( \tau _{*}\right)
\int_0^{t_\epsilon -\tau _{*}}\exp \left( dH_0t_1\right) \left( \sinh 2\xi
_{*}\left( t_\epsilon -t_1\right) -2\xi _{*}\left( t_\epsilon -t_1\right)
\right) dt_1  \nonumber \\
\ &=&C\frac{\pi \gamma }{H_0^4}a^3\left( \tau _{*}\right) \exp \left[
dH_0\left( t_\epsilon -\tau _{*}\right) \right] \int_0^{\xi _{\max }}\exp
\left( -dH_0t_p\left( \xi \right) \right) \left( \sinh 2\xi -2\xi \right) 
\frac{H_0dt_p}{d\xi }d\xi .  \label{VStarInt1}
\end{eqnarray}
Here, $\xi _{*}\left( t\right) $ is the solution of (\ref{EquForXi}) with $t$
instead of $t_\epsilon $ at the right hand side, and $t_p\left( \xi \right) $
is the inverse function, 
\begin{equation}
t_p\left( \xi \right) =\frac 2{H_0}\ln \cosh \frac \xi 2=\frac \xi {H_0}-%
\frac 2{H_0}\ln \frac 2{1+e^{-\xi }}.  \label{TPSimple}
\end{equation}
The time $t_p\left( \xi \right) +\tau _{*}$ is the proper time until
thermalization along a co-moving geodesic that thermalizes at a given value
of $\xi $; the formula (\ref{TPSimple}) was derived for the simple case of
unchanging expansion rate $H_0$. In Appendix B we find, for the case of $%
H_1/H_0\equiv h\neq 1$, an expression for $t_p\left( \xi \right) $ similar
to (\ref{TPSimple}): 
\begin{equation}
H_0t_p\left( \xi \right) =\xi -\frac{1+h}h\ln \frac{1+h}{1+he^{-\xi }}.
\label{TPFull}
\end{equation}
This coincides with (\ref{TPSimple}) for $h=1$.

The integration in (\ref{VStarInt1}) is performed up to $\xi _{\max }\equiv
\xi _{*}\left( t_\epsilon -\tau _{*}\right) $. Since $d\approx 3$, and $%
H_0t_p\left( \xi \right) \sim \xi $ for $\xi \gg 1$, the integrand of (\ref
{VStarInt1}) decays exponentially at large $\xi $, so the precise value of $%
\xi _{\max }$ is unimportant, and we can take the limit $\xi _{\max
}\rightarrow \infty $. The resulting integral with $t_p\left( \xi \right) $
given by (\ref{TPFull}) depends only on $h=H_1/H_0$ and can be expanded in $%
\left( 3-d\right) $ as 
\begin{equation}
\int_0^\infty \frac{\sinh 2\xi -2\xi }{\exp \left( dH_0t_p\left( \xi \right)
\right) }H_0\frac{dt_p}{d\xi }d\xi =f\left( h\right) +O\left( 3-d\right) ,
\label{XiInt}
\end{equation}
where the function $f\left( h\right) $ can be approximated \cite{FnFunction}
within an error of $2\%$ by 
\begin{equation}
f\left( h\right) \approx \frac{15+17h}9.  \label{FHFit}
\end{equation}

Keeping only the leading term of the expansion in $\left( 3-d\right) $, Eq.\
(\ref{VStarInt1}) for the thermalized volume becomes 
\begin{equation}
{\cal V}_{*}=Cf\left( h\right) \frac{\pi \gamma }{H_0^4}\exp \left(
dH_0\left( t_\epsilon -\tau _{*}\right) \right) a_{*}^3,  \label{VStarAsymp}
\end{equation}
where $a_{*}\equiv a\left( \tau _{*}\right) $.

The cutoff time $t_\epsilon $ is found from (\ref{CutoffCond}), 
\begin{equation}
\exp \left[ -\left( 3-d\right) H_0\left( t_\epsilon -\tau _{*}\right)
\right] =\epsilon ,  \label{CutoffCond1}
\end{equation}
and we obtain, after substituting in (\ref{VStarAsymp}) and simplifying, 
\begin{equation}
{\cal V}_{*}=Cf\left( h\right) \frac{\pi \gamma }{H_0^4}\epsilon ^{-\frac d{%
3-d}}a_{*}^3.  \label{VStarReg}
\end{equation}

The expression (\ref{VStarReg}) for the thermalized volume holds if there is
only one type of bubbles. In the case of several bubble types, the argument
above is modified in the following points: (i) the nucleation rates $\gamma
^{(j)}$ , the thermalization times $\tau _{*}^{(j)}$ and the volume
expansion factors $a_{*}^{(j)}$ are specific for the $j$-th type of bubbles;
(ii) the fractal structure of the region outside bubbles is affected by
nucleation of bubbles of all types; the corresponding fractal dimension $%
\tilde d$ is 
\begin{equation}
\tilde d=3-\frac{4\pi }{3H_0^4}\sum_j\gamma ^{(j)}\equiv 3-\frac{4\pi }{%
3H_0^4}\tilde \gamma ;  \label{FracDimJ}
\end{equation}
(iii) the cutoff condition (\ref{CutoffCond1}) is modified for bubbles of
type $j$ to 
\begin{equation}
P_{\text{outside}}\left( t_\epsilon ^{(j)}-\tau _{*}^{(j)}\right) =\exp
\left( -\left( 3-\tilde d\right) H_0\left( t_\epsilon ^{(j)}-\tau
_{*}^{(j)}\right) \right) =\epsilon .  \label{CutoffCondJ}
\end{equation}
The motivation for (\ref{CutoffCondJ}) is as follows. The cutoff procedure
for bubbles of type $j$ sets the cutoff time $t_\epsilon ^{(j)}$ at which a
fraction $\epsilon $ of all co-moving volume that will eventually thermalize
in bubbles of type $j$, is still not thermalized. Since bubbles nucleate at
time-independent rates $\gamma ^{(j)}$ per spacetime volume, the probability
for a given observer outside any bubbles to thermalize in a bubble of type $%
j $ is at all times proportional to $\gamma ^{(j)}$. Therefore, at any time $%
t$, a fraction $\gamma ^{(j)}/\tilde \gamma $ of the co-moving volume that
is outside bubbles at time $t$, and the same fraction $\gamma ^{(j)}/\tilde 
\gamma $ of the total co-moving volume, will eventually thermalize in
bubbles of type $j$. According to Eq.\ (\ref{ProbOutside}), a fraction $\exp
\left( -\tilde \gamma V_{lc}\left( t\right) \right) $ of all co-moving
volume is still outside bubbles at a time $t$; then also a fraction $\exp
\left( -\tilde \gamma V_{lc}\left( t\right) \right) $ of the co-moving
volume that is to thermalize in bubbles of type $j$, is outside bubbles at
time $t$, and this holds independent of $j$. Hence the cutoff condition (\ref
{CutoffCond1}) is only modified for a given type $j$ in its dependence on $d$
and $\tau _{*}$, as written in (\ref{CutoffCondJ}).

The regularized thermalized volume ${\cal V}_{*}^{(j)}$ corresponding to
bubbles of type $j$ becomes 
\begin{equation}
{\cal V}_{*}^{(j)}=Cf\left( h^{(j)}\right) \frac{\pi \gamma ^{(j)}}{H_0^4}%
\epsilon ^{-\frac{\tilde d}{3-\tilde d}}\left[ a_{*}^{(j)}\right] ^3.
\label{VStarRegJ}
\end{equation}
The ratio of volumes in bubbles of types, e.g., $1$ and $2$ is 
\begin{equation}
\frac{{\cal V}_{*}^{(1)}}{{\cal V}_{*}^{(2)}}\approx \frac{\gamma ^{(1)}}{%
\gamma ^{(2)}}\left[ \frac{a_{*}^{(1)}}{a_{*}^{(2)}}\right] ^3\frac{f\left(
h^{\left( 1\right) }\right) }{f\left( h^{\left( 2\right) }\right) }.
\label{VStarRatio}
\end{equation}
Since the ratio is independent of $\epsilon $, the ratios of thermalized
volumes in bubbles of different types are directly given by Eq.\ (\ref
{VStarRatio}) \cite{OtherRegV}.

\section{Arbitrary time variables\label{TIMES}}

We consider now a different choice of time variable $\bar t$ related to the
proper time $t$, along a geodesic $r=r_0$, by: 
\begin{equation}
d\bar t=T\left( H\left( t,r_0\right) \right) dt,  \label{NewTime}
\end{equation}
where $T\left( H\right) $ is an arbitrary (positive) function. Such a
relation will, for instance, describe the proper time ($T\equiv 1$) and the
``scale factor'' time ($T\left( H\right) =H$). We can always normalize $\bar 
t$ so that $T\left( H_0\right) =1$. Then, the new time variable $\bar t$
will be identical to $t$ in de Sitter regions where $H=H_0$. However, inside
bubbles the time variable will be significantly changed. In this Section, we
will modify the calculations of the preceding sections to accommodate the
new time variable and show that the result (\ref{VStarRatio}) is independent
of the choice of $T\left( H\right) $.

As in Sec.\ \ref{GEOM}, we calculate the time along a co-moving de Sitter
geodesic by matching it with a Robertson-Walker geodesic at a time $\tau _0$%
. The thermalization time (\ref{TauStar}) is then modified to 
\begin{equation}
\bar \tau _{*}=\tau _0+\int_{\tau _0}^{\tau _{*}}T\left( H\left( \tau
\right) \right) d\tau .
\end{equation}
The calculations of the co-moving and physical volumes outside of bubbles (%
\ref{ProbOutside})$-$(\ref{VInf}) and of the number of nucleated bubbles (%
\ref{NBubbles}) concern only the de Sitter region, therefore for the new
time variable the same expressions hold, and the fractal dimension $d$ is
unchanged. Equation (\ref{CutoffCond}) for the cutoff $t_\epsilon $ is
modified to 
\begin{equation}
P_{outside}\left( \bar t_\epsilon -\bar \tau _{*}\right) =\epsilon .
\end{equation}
In the calculation of the thermalized volume (\ref{ThermVol}), the
integration is performed on the thermalization surface that does not depend
on time parametrization, so the result (\ref{ThermVol}) holds. The spatial
cutoff $\xi _{*}$ becomes 
\begin{equation}
\bar \xi _{*}=2\cosh ^{-1}\exp \frac{H_0\left( \bar t_\epsilon -\bar \tau
_{*}\right) }2.
\end{equation}
The regularized thermalized volume is found analogously to (\ref{VStarInt1}%
), except that the integration is done over the time of bubble nucleation $%
\bar t_1$ in the new time parametrization. The calculations are identical,
except for the changed values of $\bar \tau _{*}$, and the results (\ref
{VStarRegJ}), (\ref{VStarRatio}) depend on $\bar \tau _{*}$ only through
invariant factors $a_{*}\left( \bar \tau _{*}\right) $ given by 
\begin{equation}
a_{*}\left( \bar \tau _{*}\right) =\exp \int_0^{\bar \tau _{*}}H\left( \bar 
\tau \right) d\bar \tau =\exp \left[ -8\pi \int_{\varphi _0}^{\varphi _{*}}%
\frac{V\left( \varphi \right) }{V^{\prime }\left( \varphi \right) }d\varphi
\right] ,  \label{AStar}
\end{equation}
where $\varphi _0$ and $\varphi _{*}$ are appropriate initial and final
field values. We conclude that the regularized probability ratios (\ref
{VStarRatio}) are independent of time parametrization.

\section{The Linde-Mezhlumian model\label{OMEGA}}

\begin{figure}[tbh]
\epsfysize 5 cm \epsffile{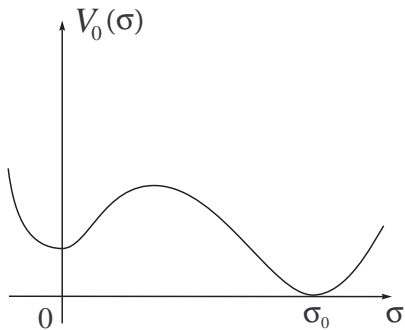}
\caption{The shape of the potential $V_0(\sigma )$ in Eq.\ (\ref{HybridPot}%
). }
\label{BubFig3}
\end{figure}

Linde and Mezhlumian \cite{LindeO1} considered a model of hybrid inflation
in which homogeneous open universes with different values of $\Omega <1$ are
created via bubble nucleation. In that model, two scalar fields $\sigma $
and $\phi $ evolve in an effective potential of the form 
\begin{equation}
V\left( \sigma ,\phi \right) =V_0\left( \sigma \right) +\sigma ^2V_1\left(
\phi \right) ,  \label{HybridPot}
\end{equation}
where the potential $V_0\left( \sigma \right) $ has two minima corresponding
to the false and true vacua, respectively (Fig.\ \ref{BubFig3}), and $%
V_1\left( \phi \right) $ is some potential with a slow-roll region suitable
for ``chaotic'' or ``new'' inflation. While the field $\sigma $ stays in the
false vacuum ($\sigma =0$), the potential for $\phi $ is flat, and quantum
fluctuations smooth out the distribution of $\phi $ to almost uniform (up to
corrections due to tunneling, see below). To make this distribution
normalizable, we shall assume that $\phi $ is a cyclic variable and identify 
$\phi =\phi _C$ with $\phi =0$. The field $\sigma $ has a small probability
to tunnel to the true vacuum through the formation of bubbles which will
have a continuous spectrum of values of $\phi $. Inside a bubble, the
potential becomes $\phi $-dependent and the field $\phi $ starts evolving
from its initial value $\phi _0$ until thermalization in the global minimum
of $V\left( \sigma ,\phi \right) $. Depending on the initial value $\phi _0$%
, the bubbles will undergo different amounts of inflation and, therefore,
will have different values of $\Omega $. We shall apply the results of Sec.\ 
\ref{PROB} to calculate the probability distribution for $\Omega $ in this
ensemble of bubbles, for a particular family (\ref{HybridPot})\ of
potentials $V\left( \sigma ,\phi \right) $. For sufficiently large values of 
$V_1\left( \phi \right) $ the tunneling is absent, since the $\sigma ^2$
term raises the true vacuum energy above that of the false vacuum. We can
choose the potential $V_1\left( \phi \right) $ so that tunneling is allowed
only for values of $\phi $ satisfying (\ref{SmallFluct}), and thus quantum
fluctuations of $\phi $ will not be dynamically important inside the
bubbles. We shall also assume that the value of $\phi $ does not change
appreciably during tunneling.

The type of bubble is now characterized by a continuous parameter $\phi _0$,
the value of $\phi $ at tunneling. To apply the result of Sec.\ \ref{PROB},
we need to supply a measure in the parameter space, i.e. a weight for the
bubbles with $\phi $ in the interval $\left( \phi _0,\phi _0+d\phi _0\right) 
$. The situation differs from Sec.\ \ref{PROB} also in that the nucleation
of bubbles of different types occurs in different regions of space. To
account for this, we describe the inflating regions of false vacuum by a
stationary solution of the diffusion equation for the volume $P\left( \phi
_0,t\right) d\phi _0$ of regions occupied by the field $\phi $ in the
interval $\left( \phi _0,\phi _0+d\phi _0\right) $ at time $t$ \cite
{Diffusion}. The diffusion equation is modified to include a ``decay'' term
for bubble nucleation: 
\begin{equation}
\frac \partial {\partial t}P\left( \phi ,t\right) =\frac \partial {\partial
\phi }\left[ \frac 1{8\pi ^2}H^{3/2}\frac \partial {\partial \phi }\left(
H^{3/2}P\left( \phi ,t\right) \right) -\frac{H^{\prime }}{4\pi }P\left( \phi
,t\right) \right] +\left( 3H-\frac{4\pi \gamma \left( \phi \right) }{3H^3}%
\right) P\left( \phi ,t\right) .  \label{DiffEqu}
\end{equation}
Here, $H\left( \phi \right) $ is the expansion rate in the false vacuum, and 
$\gamma \left( \phi \right) $ is the $\phi $-dependent tunneling rate. For
the potential (\ref{HybridPot}), which is our concern here, $H\left( \phi
\right) =H_0=const$. The stationary solution of (\ref{DiffEqu})\ can be
written as 
\begin{equation}
P\left( \phi ,t\right) =P_0\left( \phi \right) \exp \left( dH_0t\right) ,
\end{equation}
where $P_0\left( \phi \right) $ is the highest eigenvalue solution of the
stationary diffusion equation 
\begin{equation}
\frac{H_0^2}{8\pi ^2}\frac{\partial ^2P_0}{\partial \phi ^2}+\left( 3-\frac{%
4\pi \gamma \left( \phi \right) }{3H_0^4}\right) P_0=dP_0  \label{DiffEquS}
\end{equation}
with periodic boundary conditions, and $d$ is the corresponding eigenvalue.
According to Eqs.\ (\ref{NBubbles})$-$(\ref{ThermVolReg}), the resulting
thermalized volume in bubbles of a given type is proportional to the volume
of the regions of false vacuum in which bubbles of that type can nucleate.
The latter volume is proportional to $P_0\left( \phi _0\right) d\phi _0$.
Therefore, the probabilities of Sec.\ \ref{PROB} should be weighted with $%
P_0\left( \phi _0\right) d\phi _0$.

By integrating Eq.\ (\ref{DiffEquS}) over $\phi $, we obtain an expression
for $d$: 
\begin{equation}
d=3-\frac{4\pi }{3H_0^4}\frac{\int_0^{\phi _C}\gamma \left( \phi \right)
P_0\left( \phi \right) d\phi }{\int_0^{\phi _C}P_0\left( \phi \right) d\phi }%
.
\end{equation}
Since the tunneling rate $\gamma $ is small, we can approximate the solution
of (\ref{DiffEquS}) by a constant function, and then the eigenvalue $d$ is
given by the formula similar to (\ref{FracDimJ}): 
\begin{equation}
d\approx 3-\frac{4\pi }{3H_0^4}\frac 1{\phi _C}\int_0^{\phi _C}\gamma \left(
\phi \right) d\phi .  \label{DSimple}
\end{equation}

According to (\ref{VStarRatio}), the probability distribution depends on $%
\phi _0$ through the nucleation rate $\gamma \left( \phi _0\right) $, the
expansion factor $a_{*}\left( \phi _0\right) $, and the factor $f\left[
H\left( \phi _0\right) /H_0\right] \equiv f\left( \phi _0\right) $ which
describes the effect of a different expansion rate $H_1=H\left( \phi
_0\right) $ in the bubble interior after nucleation. The nucleation rate per
unit spacetime volume is estimated \cite{Turok1} using the Euclidean O$%
\left( 4\right) $-symmetric instanton solution $\sigma \left( r\right) $ for
the field $\sigma $ coupled to gravity: 
\begin{equation}
\gamma \left( \phi _0\right) =A\left( \phi _0\right) \exp \left( -S_E\left(
\phi _0\right) \right) ,
\end{equation}
where $S_E\left( \phi _0\right) $ is the instanton action and $A\left( \phi
_0\right) $ is the prefactor which we assume to be a slowly-varying function
of $\phi _0$. The regularized probability of being in bubbles that tunneled
with $\phi =\phi _0$ is then expressed, with a suitable normalization
constant $N$, as 
\begin{equation}
d{\cal P}\left( \phi _0\right) =N\nu _{\text{civ}}\left( \phi _0\right) d%
\tilde {{\cal P}}\left( \phi _0\right) ,  \label{ProbWN}
\end{equation}
where we have separated the distribution $d\tilde {{\cal P}}\left( \phi
_0\right) $ due to the thermalized volume, 
\begin{equation}
d\tilde {{\cal P}}\left( \phi _0\right) =A\left( \phi _0\right) \exp \left(
-S_E\left( \phi _0\right) \right) a_{*}^3\left( \phi _0\right) f\left( \phi
_0\right) P_0\left( \phi _0\right) d\phi _0,  \label{ProbVar}
\end{equation}
from the ``human factor'' $\nu _{\text{civ}}\left( \phi _0\right) $
introduced in Eq.\ (\ref{CivRatio}). We will now concentrate on the above
distribution, whereas the effect of the factor $\nu _{\text{civ}}\left( \phi
_0\right) $ will be discussed in the next Section. We shall be interested in
the leading (exponential) dependence on $\phi _0$ in Eq.\ (\ref{ProbVar})
and shall therefore approximate the factors $A\left( \phi _0\right) $, $%
f\left( \phi _0\right) $ and $P_0\left( \phi _0\right) $ by a constant.

The expansion factor at thermalization $a_{*}\left( \phi _0\right) $ for a
bubble formed at the value $\phi =\phi _0$ is determined by (\ref{AStar}), 
\begin{equation}
a_{*}\left( \phi _0\right) =\exp \left[ -8\pi \int_{\phi _0}^{\phi _{*}}%
\frac{V\left( \sigma _0,\phi \right) }{V_\phi ^{\prime }\left( \sigma
_0,\phi \right) }d\phi \right] ,  \label{AStar1}
\end{equation}
where the value $\phi _{*}$ corresponds to the end of slow roll and is
defined by 
\begin{equation}
\left. \frac{V^{\prime }}{4\pi V}\right| _{\phi =\phi _{*}}\simeq 1.
\end{equation}

Eqs.\ (\ref{ProbVar})$-$(\ref{AStar1}) give the probability distribution for
the value $\phi _0$ at which tunneling of the field $\sigma $ occurs. To
obtain a probability distribution for $\Omega $, we need to find the present
value of $\Omega $ as a function of $\phi _0$. As outlined in \cite{Turok1},
we can relate $\Omega $ to the expansion factor $a_{*}\left( \phi _0\right) $
given by (\ref{AStar1}): 
\begin{equation}
\Omega \left( \phi _0\right) =\left( 1+\frac B{a_{*}^2\left( \phi _0\right) }%
\right) ^{-1},\quad B\equiv \left( \frac{T_{th}}{T_{eq}}\right) ^2\frac{%
T_{eq}}{T_{CMB}},  \label{Omega}
\end{equation}
where $T_{th}$ is the thermalization temperature, $T_{CMB}$ is the cosmic
microwave background temperature at present, and $T_{eq}$ is the temperature
at equal matter and radiation density. Depending on $T_{th}$, the value of $%
B $ is $\sim 10^{25}-10^{50}$. A higher value of $V_1\left( \phi _0\right) $
corresponds to longer inflation and a larger expansion factor $a_{*}$, and
therefore to a value of $\Omega $ closer to $1$.

To calculate $d\tilde {{\cal P}}\left( \Omega \right) /d\Omega $, we choose
a potential that in the range of $\phi $ where tunneling is allowed is given
by 
\begin{equation}
V\left( \sigma ,\phi \right) =V_0\left( \sigma \right) +\frac g2\phi
^2\sigma ^2,  \label{HybridPotVar}
\end{equation}
where $V_0\left( \sigma \right) $ still has the shape shown in Fig.\ \ref
{BubFig3}. A similar potential was also considered in \cite{LindeO1}. Note
that the slow roll condition (\ref{SlowRoll}) requires $\phi \gg 1$. To
facilitate the calculation of the instanton action $S_E\left( \phi _0\right) 
$, we shall choose $V_0\left( \sigma \right) $ to be quartic in $\sigma $: 
\begin{equation}
V_0\left( \sigma \right) =\lambda \sigma ^4-b_1\sigma ^3+b_2\sigma ^2+const.
\label{QuarticV}
\end{equation}
The constant is chosen so that the true vacuum energy is zero, giving a
vanishing cosmological constant. Since we assumed that the bubble size is
small on the horizon scale, we can disregard the effect of gravity and treat
the instanton as in flat space. The calculation of the instanton action in
flat space for general quartic potentials of the form (\ref{QuarticV}) was
performed semi-analytically in \cite{Adams}, and we shall use the result
obtained there, 
\begin{equation}
S_E\left( \phi _0\right) =\frac{\pi ^2}{3\lambda }\frac{\alpha _1\delta
+\alpha _2\delta ^2+\alpha _3\delta ^3}{\left( 2-\delta \right) ^3},
\label{QuarticSE}
\end{equation}
where $\alpha _1=13.832$, $\alpha _2=-10.819$, $\alpha _3=2.0765$ and the
dimensionless parameter $\delta \left( \phi _0\right) $ is defined, in terms
of the parameters of the potential (\ref{QuarticV}), by 
\begin{equation}
\delta \left( \phi _0\right) \equiv 8\lambda \frac{b_2+\frac g2\phi _0^2}{%
b_1^2}.
\end{equation}
The allowed range of $\delta $ is from its minimum value $\delta _{\min
}=8\lambda b_2/b_1^2$ to $2$, where $\delta =2$ corresponds to the maximum
value of $\phi $ at which tunneling can still occur:, 
\begin{equation}
\phi _{\max }^2=\frac 2g\left( \frac{b_1^2}{4\lambda }-b_2\right) =\frac{%
b_1^2}{4\lambda g}\left( 2-\delta _{\text{min}}\right) .  \label{PhiMax}
\end{equation}

The thin wall approximation, valid when the minima of the potential (\ref
{QuarticV}) are almost degenerate, corresponds to $\delta _{\min }\approx 2$%
, and then $\alpha _1\delta +\alpha _2\delta ^2+\alpha _3\delta ^3\approx 1$
for $\delta _{\min }<\delta <2$. A generic choice of parameters $\lambda $, $%
b_1$ and $b_2$, such as $b_1\sim \lambda \sigma _0$, $b_2\sim \lambda \sigma
_0^2$, will give $\delta _{\min }\sim 1$. Then, the expression $\alpha
_1\delta +\alpha _2\delta ^2+\alpha _3\delta ^3$ in (\ref{QuarticSE}) is
also of order $1$ for the allowed range of $\delta $. Accordingly, we will
disregard this expression below.

Using the potential (\ref{HybridPotVar})$-$(\ref{QuarticV}), we can
calculate $a_{*}\left( \phi _0\right) $: 
\begin{equation}
a_{*}\left( \phi _0\right) =\exp \left[ -8\pi \int_{\phi _0}^{\phi _{*}}%
\frac{V\left( \sigma _0,\phi \right) }{V_\phi ^{\prime }\left( \sigma
_0,\phi \right) }d\phi \right] =\exp \left[ 8\pi \int_{\phi _{*}}^{\phi _0}%
\frac{\frac g2\phi ^2\sigma _0^2+V_0\left( \sigma _0\right) }{\frac d{d\phi }%
\left( \frac g2\phi ^2\sigma _0^2+V_0\left( \sigma _0\right) \right) }d\phi
\right] .  \label{HybridA}
\end{equation}
Since the potential (\ref{HybridPotVar}) includes interaction between $%
\sigma $ and $\phi $, the value of the true vacuum $\sigma _0$ will be
slightly $\phi _0$-dependent, $\sigma _0=\sigma _0\left( \phi _0\right) $,
and as the field $\phi $ slowly evolves toward $\phi =\phi _{*}$, the field $%
\sigma $ will follow the shifting position of the minimum $\sigma _0\left(
\phi _0\right) $. Without the dependence of $\sigma _0$ on $\phi $, the
expansion factor would be 
\begin{equation}
a_{*}\left( \phi _0\right) =\exp \left[ 8\pi \int_{\phi _{*}}^{\phi _0}\frac{%
\frac g2\phi ^2\sigma _0^2}{g\phi \sigma _0^2}d\phi \right] \approx \exp
\left( 2\pi \phi _0^2-2\pi \phi _{*}^2\right) .  \label{HybridASA}
\end{equation}
The exact expression for $a_{*}\left( \phi _0\right) $ contains a correction
to (\ref{HybridASA}), 
\begin{equation}
a_{*}\left( \phi _0\right) =\exp \left[ 2\pi \left( \phi _0^2-\phi
_{*}^2\right) \left( 1+F\left( \frac{\phi _0}{\phi _{\max }}\right) \right)
\right] ,  \label{HybridAS}
\end{equation}
where the function $F$ behaves as $F\left( x\right) \sim x^2$ at small $x$
and $F\left( 1\right) \lesssim 1$ (the explicit form of $F$ is unimportant).

Eq.\ (\ref{Omega}) for $\Omega \left( \phi _0\right) $ becomes 
\begin{equation}
\Omega \left( \phi _0\right) =\left[ 1+B\exp \left( -4\pi \left( 1+F\right)
\left( \phi _0^2-\phi _{*}^2\right) \right) \right] ^{-1}.  \label{OmegaPhi0}
\end{equation}
Assuming that $\ln B\gg 1$, one can see that $\Omega \left( \phi _0\right) $
changes very quickly from $0$ to $1$ in a narrow region of relative width $%
\Delta \phi /\phi \sim \left( \ln B\right) ^{-1}$ around $\phi =\phi _1$,
where $\phi _1=\sqrt{\left( 1/4\pi \right) \ln B+\phi _{*}^2}$. For a
typical value of $\ln B\sim 100$, one obtains $\phi _1\sim 3$. Note that for 
$\phi _0<1$ the slow roll approximation is not valid and Eqs.\ (\ref{HybridA}%
)$-$(\ref{OmegaPhi0}) are not applicable; we shall only consider the
distribution (\ref{ProbVar}) for $\phi _0>\phi _{*}$, where $\phi _{*}\sim 1$%
. Correspondingly, the range of $\Omega $ is from $\Omega \left( \phi
_{*}\right) \sim B^{-1}\approx 0$ to $\Omega _{\max }\equiv \Omega \left(
\phi _{\max }\right) $. The maximum value of $\Omega $ is 
\begin{equation}
\Omega _{\max }=\left[ 1+B\exp \left( -4\pi \left( 1+F\left( 1\right)
\right) \phi _{\max }^2\right) \right] ^{-1}.
\end{equation}
Generically, $\phi _{\max }\gg 1$ and $\Omega _{\max }$ is very close to $1$.

Combining (\ref{ProbVar}), (\ref{QuarticSE}), and (\ref{HybridAS}), we
obtain the leading exponential dependence of the distribution (\ref{ProbWN})
on $\phi _0$: 
\begin{equation}
\frac{d\tilde {{\cal P}}\left( \phi _0\right) }{d\phi _0}\propto \exp \left(
-\frac{\pi ^2}{3\lambda }\left( 2-\delta \left( \phi _0\right) \right)
^{-3}+6\pi \left( 1+F\right) \left( \phi _0^2-\phi _{*}^2\right) \right) .
\label{HybridProb}
\end{equation}

\section{Probability distribution for $\Omega \label{CIV}$}

\begin{figure}[tbh]
\epsfysize 5 cm \epsffile{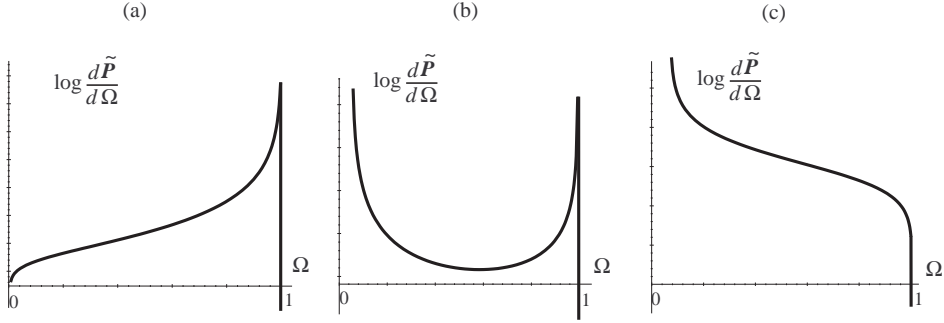}
\caption{Probability distribution $d\tilde {{\cal {P}}}(\Omega )/d\Omega $
with $\phi _{\max }^2/\phi _{*}^2=100$, shown logarithmically up to a
normalization. (a): $\mu =0.01$. The peak at $\Omega \approx 1$ is extremely
sharp; the ratio of the values at $\Omega =.99$ and at $\Omega =0$ is $\sim
\exp 20$, while the peak value differs from that at $\Omega =0$ by a factor
of $\sim \exp \left( 25000\right) $. (b): $\mu =0.5$, there are two local
maxima near $\Omega =0$ and $\Omega =1$. (c): $\mu =2$. The function
monotonically decreases. The maximum value at $\Omega =0$ differs from a
typical intermediate value ($\Omega \sim 1/2$) by a factor of $\sim \exp
\left( 60\right) $. }
\label{BubFig4}
\end{figure}

To obtain the probability distribution for $\Omega =\Omega \left( \phi
_0\right) $, we need to transform $d\tilde {{\cal P}}\left( \phi _0\right) $
to the new variable $\Omega $ via 
\begin{equation}
d\tilde {{\cal P}}\left( \Omega \right) =\frac{d\tilde {{\cal P}}\left( \phi
_0\right) }{d\phi _0}\left( \frac{d\Omega }{d\phi _0}\right) ^{-1}d\Omega .
\label{ProbTilde}
\end{equation}
Expressed as a function of $\phi _0$, this distribution is 
\begin{equation}
\frac{d\tilde {{\cal P}}\left( \Omega \right) }{d\Omega }=\exp \left( -\frac{%
\pi ^2\left( b_1/4\lambda g\right) ^3}{3\lambda \left( \phi _{\max }^2-\phi
_0^2\right) ^3}+\pi \left( 10+6F\right) \left( \phi _0^2-\phi _{*}^2\right)
\right) \frac 1{8\pi B\phi _0\Omega ^2}.  \label{ProbOmega}
\end{equation}
Since most of the range of $\Omega $ (except a narrow region around $\Omega
=1$) corresponds to $\phi _0\ll \phi _{\max }$, we can expand (\ref
{ProbOmega}) in $\phi _0^2/\phi _{\max }^2$ and obtain an approximate
power-law dependence 
\begin{equation}
\frac{d\tilde {{\cal P}}\left( \Omega \right) }{d\Omega }\propto \Omega
^{1/2-3\mu }\left( 1-\Omega \right) ^{3\mu -5/2},  \label{PowerLaw}
\end{equation}
where we have defined the dimensionless parameter $\mu $ by 
\begin{equation}
\mu =\frac \pi {12\lambda \phi _{\max }^2\left( 2-\delta _{\min }\right) ^3}.
\label{Mu}
\end{equation}
For $\Omega $ very close to $1$, the right hand side of (\ref{ProbOmega}) is
dominated by the first term in the exponential, which makes it rapidly drop
to $0$. Depending on the value of $\mu $, there are three distinct behaviors
of $d\tilde {{\cal P}}/d\Omega $ (Fig.\ \ref{BubFig4}). In the first case, $%
\mu <1/6$, the function monotonically grows with $\Omega $ until it peaks at 
$\Omega =\Omega _{\text{peak}}\approx \Omega _{\max }\approx 1$ and very
rapidly falls off to $0$ for $\Omega >\Omega _{\text{peak}}$. The second
case occurs for $\mu >5/6$; the distribution (\ref{ProbTilde}) monotonically
decreases with $\Omega $ and its maximum is at the lower boundary $\Omega =0$%
. Lastly, in the third case, with $1/6<\mu <5/6$, the distribution (\ref
{ProbOmega}) decreases from a local maximum at $\Omega =0$ and then
increases to another local maximum at $\Omega =\Omega _{\text{peak}}\approx 1
$ (Fig. \ref{BubFig6}). To determine which maximum dominates the probability
distribution, we consider its approximate form (\ref{PowerLaw}). If $\mu <1/2
$, the second exponent in Eq.\ (\ref{PowerLaw}) is smaller than the first
one, giving a stronger peak at $\Omega \approx 1$. For $\mu >1/2$, the peak
at $\Omega =0$ is stronger \cite{OtherRegP}.

Now we consider the influence of the factor $\nu _{\text{civ}}\left( \phi
_0\right) $ on the distribution (\ref{ProbOmega}). In our model, the
low-energy physics is identical in all bubbles, and therefore $\nu _{\text{%
civ}}$ is simply proportional to the number of potentially inhabitable
stellar systems. The structure formation process in different bubbles is
also essentially the same, apart from the difference in $\Omega $. The main
effect of $\Omega $ is to terminate the growth of density fluctuations at
redshift $1+z_\Omega \approx \Omega ^{-1}$ \cite{FluctO}. Assuming that the
dominant matter compontent is ``cold'', density fluctuations begin to grow
at redshift of matter and radiation equality, $1+z_{\text{eq}}\approx
2\times 10^4\Omega h^2$. With $h=0.7$, the overall growth factor is 
\begin{equation}
f\left( \Omega \right) =\frac{1+z_{\text{eq}}}{1+z_\Omega }\approx
10^4\Omega ^2,
\end{equation}
where we have assumed that $z_{eq}>z_\Omega $, that is, $\Omega >10^{-2}$.
Otherwise, there is no growth, and thus $f\left( \Omega \right) \sim 1$ for $%
\Omega \lesssim 10^{-2}$.

\begin{figure}[tbh]
\epsfysize 5 cm \epsffile{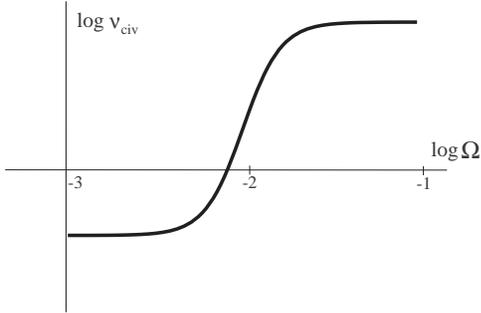}
\caption{Dependence of $\nu _{\text civ}$ on $\Omega $ for $\bar
\delta=10^{-2}$. The ratio of the
maximum and the minimum values is $\sim \exp \bar \delta ^{-2}$. The
origin on the vertical axis is arbitrary.}
\label{BubFig5}
\end{figure}

\begin{figure}[tbh]
\epsfysize 5 cm \epsffile{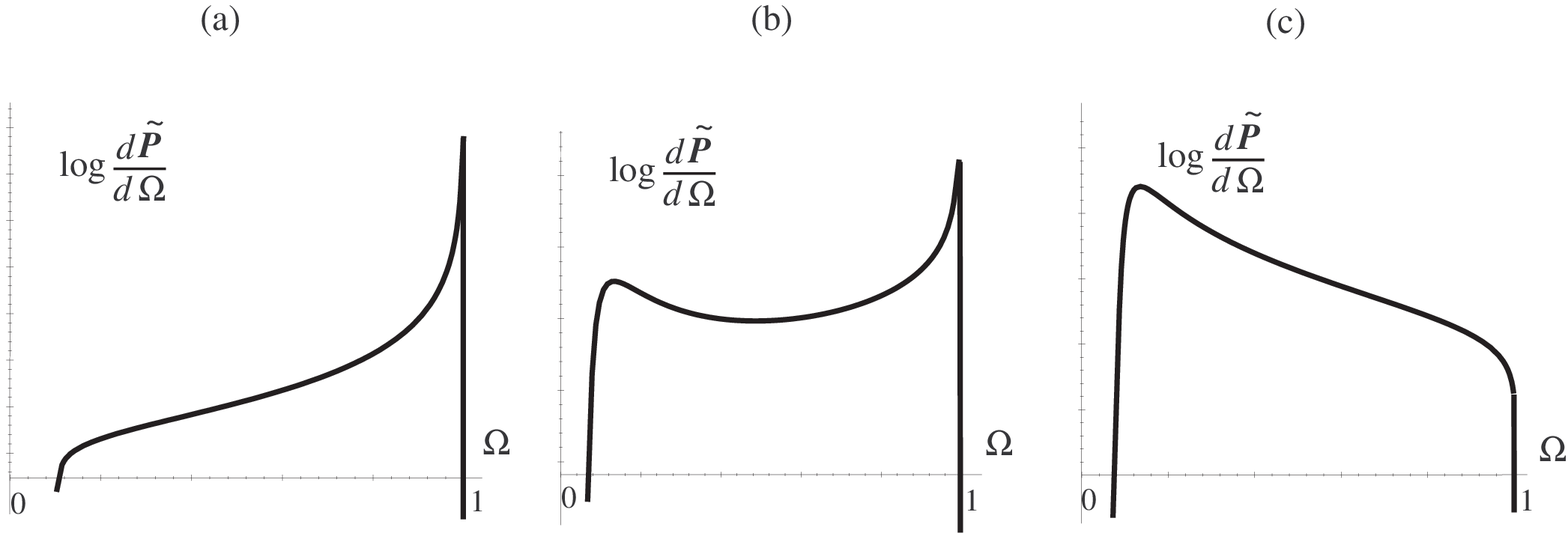}
\caption{Probability distribution $d {\cal {P}}/d\Omega $, shown
logarithmically up to a normalization, in the three cases corresponding to
Fig.\ \ref{BubFig4}a-c. The parameter values are the same as those in
Figs.\ \ref{BubFig4}, \ref{BubFig5}.}
\label{BubFig6}
\end{figure}

If density fluctuations are generated by inflation, then their initial
amplitude on each scale has a Gaussian distribution. Its rms value at
horizon crossing, $\left( \delta \rho /\rho \right) _{\text{rms}}\equiv \bar 
\delta $, is determined by the shape of the potential $V\left( \varphi
,\sigma \right) $ and is approximately scale-independent on astrophysically
relevant scales. In bubbles with values of $\Omega $ such that $f\left(
\Omega \right) \bar \delta >1$, most of the matter is captured into bound
objects, and $\nu _{\text{civ}}$ is essentially independent of $\Omega $. On
the other hand, if $f\left( \Omega \right) \bar \delta \ll 1$, then almost
no structure is formed. In this case, bound objects are formed only in the
rare regions where $\delta \rho /\rho $ exceeds the rms value $\bar \delta $
by a factor $\gtrsim \left[ f\left( \Omega \right) \bar \delta \right]
^{-1}\gg 1$. Hence, we expect that in the range $10^{-2}<\Omega \ll \bar 
\Omega $, 
\begin{equation}
\nu _{\text{civ}}\left( \Omega \right) \propto \exp \left( -\kappa \Omega
^{-4}\right) ,
\end{equation}
where $\kappa \sim 10^{-8}\bar \delta ^{-2}$ and $\bar \Omega \sim 10^{-2}%
\bar \delta ^{-1/2}$ is the solution of $f\left( \Omega \right) \bar \delta
\sim 1$. For $\Omega <10^{-2}$, $f\left( \Omega \right) \sim 1$ and we
expect $\nu _{\text{civ}}\left( \Omega \right) \propto
\exp \left( -\kappa ^{\prime }%
\bar \delta ^{-2}\right) $ with $\kappa ^{\prime }\sim 1$. The function $\nu
_{\text{civ}}\left( \Omega \right) $ is sketched in Fig.\ \ref{BubFig5} for
the full range of $\Omega $ \cite{Fn4}.

The effect of the factor $\nu _{\text{civ}}\left( \Omega \right) $ on the
probability distribution (\ref{ProbTilde}) can now be easily understood. If $%
d\tilde {{\cal P}}/d\Omega $ has a single peak near $\Omega =1$, as in Fig.\ 
\ref{BubFig4}a, then the peak position remains essentially unchanged, and
the distribution function is suppressed only for $\Omega <\bar \Omega $,
where it was already very small. The most interesting modification occurs
when there is a (local) peak at $\Omega =0$, as in Figs.\ \ref{BubFig4}b-c.
This peak is then shifted to a larger value, $\Omega _{\text{peak}}\approx 
\sqrt{2}\left( 3\mu -1/2\right) ^{-1/4}\bar \Omega \sim \bar \Omega$. (For
$\bar \delta \sim 10^{-3}$, $\bar \delta \sim 10^{-2}$, and $\bar
\delta \sim 10^{-1}$, we
obtain $\bar \Omega \sim 0.3$, $\bar \Omega \sim 0.1$, and $\bar \Omega \sim
0.03$, respectively.) In the case of $\mu >5/6$, this is the only maximum of 
$d\tilde {{\cal P}}/d\Omega $. The behavior of the full probability
distribution (\ref{ProbWN}), 
\begin{equation}
d{\cal P}\left( \Omega \right) =N\nu _{\text{civ}}\left( \Omega \right) d%
\tilde {{\cal P}}\left( \Omega \right) ,
\end{equation}
is sketched in Fig.\ \ref{BubFig6} for all three cases.

The idea that anthropic considerations make a low value of $\Omega $ very
unlikely has been previously discussed by a number of authors \cite
{Anthropic1,Anthropic3}; however, to our knowledge, no attempt has been made
to make this argument quantitative. A similar approach to the cosmological
constant has been developed in Ref.\ \cite{Cosm1,Cosm2,Predictions,Cosm3}.

\section{Conclusions\label{CONCL}}

In this paper, we have considered scenarios of open-universe inflation,
where the metastable false vacuum decays by quantum tunneling and forms
bubbles of different types. The interior of a bubble is observed as an open
universe, in which inflation continues until thermalization. Our goal was to
find a probability distribution for thermalization in different bubble
types, following the approach of \cite{Predictions,MakingPredictions},
according to which the probability is proportional to the number of
civilizations that will evolve in bubbles of each type. The problem of
calculating the probability splits into a calculation of the ratio of
physical volumes thermalized in different types of bubbles and of the number
of civilizations $\nu _{\text{civ}}$ that evolve per unit thermalized volume.

In Sections \ref{GEOM}$-$\ref{OMEGA} we developed a method for calculating
the volume ratios. We first considered the case of a discrete set of bubble
types and then extended the analysis to a continuous spectrum of bubbles. As
an example of the latter, we focussed on the Linde-Mezhlumian model of
hybrid inflation \cite{LindeO1} which gives rise to an ensemble of open
universes with $\Omega <1$. Since all nucleating bubbles in this model have
identical particle physics, we were able also to estimate the ``human
factor'' $\nu _{\text{civ}}\left( \Omega \right) $. We found that, depending
on the dimensionless parameter $\mu $ defined in (\ref{Mu}), the probability
distribution $d{\cal P}/d\Omega $ is peaked either at $\Omega =1$ (for $\mu
<1/6$) or at an intermediate value $\Omega =\bar \Omega $ in the range $%
0.03\lesssim \bar \Omega <1$ (for $\mu >5/6$). For $1/6<\mu <5/6$, the
distribution has local maxima at both $\Omega =\bar \Omega $ and $\Omega =1$%
; the relative magnitude of the two peaks depends on $\mu $ and on the
amplitude $\bar \delta $ of density fluctuations.

In this paper, we have only considered models in which the false-vacuum
regions inflate at a constant rate $H$. In a more general situation, the
field $\phi $ would evolve before as well as after tunneling, leading to a
slowly changing $H$. The analysis of such models would be substantially more
complicated, while the results are likely to be similar to those in simpler
models with a constant $H$.

\appendix 

\section{Proper time in the bubble interior}

Here it will be shown that a co-moving geodesic continued from the de Sitter
region to the bubble interior exponentially approaches a Robertson-Walker
(RW) stationary geodesic inside the bubble. We shall calculate the proper
time along such a geodesic and show that the approximate formula (\ref
{ProbInf}) is accurate within our assumptions.

As we noted in Sec.\ \ref{GEOM}, there is a region inside the bubble in
which the spacetime is approximately de Sitter, and the RW coordinates $%
\left( \tau ,\xi \right) $ in that region are related to de Sitter ones by (%
\ref{NewCoords}). The range of $\tau $ in that region is $\tau \ll
1/H^{\prime }\left( \varphi _0\right) $, as follows from (\ref
{TauSmallerThan}). We can use the coordinate change (\ref{NewCoords}) to
continue a co-moving geodesic $r=r_0$ from the false vacuum region to the
bubble interior (provided that the geodesic intersects the bubble, i.e. that 
$H_0r_0<1$). The resulting trajectory $\xi \left( \tau \right) $ is 
\begin{equation}
\xi \left( \tau \right) =\ln \frac{H_0r_0\cosh H_0\tau +\sqrt{\sinh
^2H_0\tau +\left( H_0r_0\right) ^2}}{\left( 1-H_0r_0\right) \sinh H_0\tau }.
\label{ComovingXi}
\end{equation}
At large values of $\tau $ such that $\exp H_0\tau \gg 1$, the trajectory (%
\ref{ComovingXi}) becomes 
\begin{equation}
\xi \left( \tau \right) =\ln \frac{1+H_0r_0}{1-H_0r_0}+\ln \left( 1+\frac{%
e^{-H_0\tau }}2\frac{\left( H_0r_0\right) ^2}{1+H_0r_0}+O\left( e^{-2H_0\tau
}\right) \right) =const+O\left( e^{-H_0\tau }\right) ,
\label{AsympComovingXi}
\end{equation}
i.e. it is exponentially close to the co-moving geodesic line $\xi =const$
in the RW region.

We see from (\ref{SlowRoll}), (\ref{TauSmallerThan}) that there is a range
of $\tau $ such that 
\begin{equation}
1\ll H_0\tau \ll \frac{H_0}{H^{\prime }\left( \varphi _0\right) },
\label{RangeOfTau}
\end{equation}
and in this range the co-moving world-lines, $r=r_0$, continued from the
region outside of the bubble into the interior, become very close to the RW
co-moving world-lines, $\xi =\xi _0$, while the spacetime is still
sufficiently close to de Sitter. At times $\tau $ satisfying (\ref
{RangeOfTau}), the proper time along $r=r_0$ becomes exponentially close to $%
\tau $, as shown by (\ref{TTau}).

Now we will consider Eq.\ (\ref{ProbInf}) which was based on the assumption
that the time interval between crossing the bubble boundary and
thermalization is equal to $\tau _{*}$ for all geodesics. This assumption is
not exactly true, because during the time period when the time variables $t$
and $\tau $ differ significantly, their difference depends on the spatial
coordinate $r_0$, which varies among different geodesics $r=r_0$. As a
result, the proper time interval along a geodesic between entering the
bubble and the point $\left( \tau ,\xi \right) $ differs from $\tau $ by an $%
r_0$-dependent correction $\Delta \tau $. Assume for simplicity that the
bubble is centered at $r=0$. A co-moving geodesic $r=r_0$ entered the bubble
at time $t_0$ given by 
\begin{equation}
t_0=-\left( 1/H_0\right) \ln \left( 1-H_0r_0\right) .
\end{equation}
The correction $\Delta \tau $ is then 
\begin{equation}
\Delta \tau \left( r_0\right) \equiv t\left( \tau ,\xi \left( r_0\right)
\right) -t_0-\tau =\frac 2{H_0}\ln \cosh \frac{\xi \left( r_0\right) }2-t_0=-%
\frac 1{H_0}\ln \left( 1+H_0r_0\right) .  \label{DeltaTau}
\end{equation}
The function $\Delta \tau \left( r_0\right) $ does not depend on $\tau $ and
its maximum value is $-H_0^{-1}\ln 2$ (for $H_0r_0\approx 1$).

Now we can show that the correction (\ref{DeltaTau}) does not significantly
influence Eq.\ (\ref{ProbInf}). A change in the thermalization time $\tau
_{*}$ by the correction $\Delta \tau \sim H_0^{-1}$ in (\ref{ProbInf}) would
change Eq.\ (\ref{ProbInf}) by the factor $\exp \left( -4\pi \gamma /\left(
3H_0^4\right) \right) $, which is very close to $1$ because, as we assumed
in (\ref{SmallGamma}), $\gamma /\left( 3H_0^4\right) \ll 1$. Therefore, Eq.\
(\ref{ProbInf}) is accurate within our assumptions.

\section{The case of different expansion rates $H_1\neq H_0$}

Here we present the calculations of the proper time until thermalization in
the general case when the gravitational effect of the bubble wall is not
assumed to be small and the expansion rate $H_1$ inside the bubble
significantly different from $H_0$ (presumably, $H_1<H_0$). We shall assume,
however, that the size of the nucleated bubbles is small on the horizon
scale $H_0^{-1}$.

The de Sitter spacetime is represented by the hyperboloid 
\begin{equation}
{\bf \zeta }^2+w^2-v^2=H_0^{-2}  \label{OuterHyp}
\end{equation}
embedded in a 5-dimensional space $\left( \zeta _1,\zeta _2,\zeta
_3,w,v\right) $ with Minkowskian signature. For simplicity, we treat the
bubble interior also as a de Sitter spacetime region with constant expansion
rate $H_1$. Then the bubble interior will be a piece of the hyperboloid 
\begin{equation}
{\bf \zeta }^2+\left( w-\Delta w\right) ^2-v^2=H_1^{-2}
\end{equation}
cut out by intersection with (\ref{OuterHyp}). The displacement $\Delta w$
is related to the bubble wall tension or, alternatively, to the initial
bubble size \cite{BGV}. The bubble wall is at $w=w_0=\left( 1/2\right)
\left[ \Delta w-\left( H_1^{-2}-H_0^{-2}\right) /\Delta w\right] $.

The flat RW coordinates $\left( t,{\bf x}\right) $ in the outer region are
introduced by 
\begin{mathletters}
\label{NewCoords1}
\begin{eqnarray}
H_0t &=&\ln H_0\left( w+v\right) , \\
H_0{\bf x} &=&\frac{{\bf \zeta }}{w+v}.
\end{eqnarray}
This gives the trajectory of the bubble wall in these coordinates, 
\end{mathletters}
\begin{equation}
H_0r_0\left( t\right) =\sqrt{\left( 1-e^{-H_0t}\right) ^2+2e^{-H_0t}\left(
1-H_0w_0\right) }.
\end{equation}
The assumption of small initial bubble size corresponds to $H_0w_0\approx 1$%
, which means that we can approximate the bubble wall by the lightcone $%
H_0r_{lc}\left( t\right) =1-e^{-H_0t}$. This considerably simplifies the
algebra.

Our goal is to find the proper time until thermalization along a co-moving
geodesic that starts as $r=const$ in the outer region and crosses the bubble
wall. We introduce the flat RW coordinates $\left( t_1,{\bf x}_1\right) $
also in the interior region: 
\begin{mathletters}
\label{NewCoords2}
\begin{eqnarray}
H_1t_1 &=&\ln H_1\left( w-\Delta w+v\right) , \\
H_1{\bf x}_1 &=&\frac{{\bf \zeta }}{w-\Delta w+v}.
\end{eqnarray}
The two coordinate systems are matched at the bubble wall, and the metric is
continuous across the wall. This allows us to continue the geodesic $r=r_0$
through the bubble wall by requiring that the component of its $4$-velocity
parallel to the wall be continuous. We denote by $\beta $ the $r$ component
of the initial $4$-velocity, $\beta =\frac{dr_1}{ds}$, found from this
condition. The (generally non-zero) velocity $\left. \frac{dr_1}{dt_1}%
\right| _{t_{10}}$ with which the geodesic emerges in the interior is
determined by $\beta $. A general radial geodesic in the interior de Sitter
region is described by 
\end{mathletters}
\begin{equation}
H_1r_1\left( t_1\right) =r_{10}+\sqrt{P^{-2}+e^{H_1t_{10}}}-\sqrt{%
P^{-2}+e^{H_1t_1}},  \label{Geod}
\end{equation}
where $\left( t_{10},r_{10}\right) $ is the initial point at the bubble wall
in the coordinates $\left( t_1,{\bf x}_1\right) $ and $P$ is a constant of
motion related to the initial velocity by 
\begin{equation}
P=\frac{\exp \left( 2H_1t_{10}\right) \left. \frac{dr_1}{dt_1}\right|
_{t_{10}}}{\sqrt{1-\exp \left( 2H_1t_{10}\right) \left( \left. \frac{dr_1}{%
dt_1}\right| _{t_{10}}\right) ^2}}=\beta \exp \left( H_1t_{10}\right) .
\end{equation}
The proper time $\delta t$ along this geodesic from the bubble wall crossing
until time $t_1$ is found to be 
\begin{equation}
\delta t\left( t_1\right) =H_1^{-1}\left( \sinh ^{-1}\frac{e^{H_1t_1}}P%
-\sinh ^{-1}\frac{e^{H_1t_{10}}}P\right) .
\end{equation}

The geodesic (\ref{Geod}) asymptotes to the line $r=r_a$ at large times,
where $r_a$ is given by 
\begin{equation}
r_a=r_{10}+\frac 1{H_1}\frac{\beta \exp \left( -H_1t_{10}\right) }{1+\sqrt{%
1+\beta ^2}}.
\end{equation}
As in Sec.\ \ref{GEOM}, we introduce the open RW coordinates $\left( \tau
,\xi \right) $ in the interior and match the geodesic (\ref{Geod}) with a
line $\xi =\xi _{*}$ at time $t_1\left( \tau _0,\xi _{*}\right) $ given by (%
\ref{NewCoords-t}). This enables us to find the total proper time until
thermalization as the sum of the time $t_0$ until bubble wall crossing, the
time $\delta t\left( t_1\left( \tau _0,\xi _{*}\right) \right) $ from the
wall crossing to matching with $\xi =\xi _0$, and the time $\tau _{*}-\tau
_0 $ until thermalization: 
\begin{equation}
t_{\text{total}}\left( \xi _0\right) =t_0+\delta t\left( t_1\left( \tau
_0,\xi _0\right) \right) +\tau _{*}-\tau _0.  \label{TotalTime}
\end{equation}
The trajectory (\ref{Geod}) is completely specified by its asymptotic value
of $\xi $, and we can express the parameters $\beta $, $P$, $r_1$, $t_{10}$
and $t_0$ through $\xi _0$. After some algebra, we arrive at the following
expression for the time (\ref{TotalTime}): 
\begin{equation}
t_{\text{total}}\left( \xi _0\right) =\tau _{*}+\frac 1{H_0}\left( \xi _0-%
\frac{1+h}h\ln \frac{1+h}{1+he^{-\xi _0}}\right) .
\end{equation}
For $h=1$, this reduces to the left hand side of (\ref{EquForXi}), as
expected.

In the calculation of the thermalized volume in Sec.\ \ref{PROB}, we will
use the function $t_p\equiv t_{\text{total}}\left( \xi _0\right) -\tau _{*}$%
, which has the meaning of the correction to the thermalization time: 
\begin{equation}
H_0t_p\left( \xi _0\right) =\xi _0-\frac{1+h}h\ln \frac{1+h}{1+he^{-\xi _0}}.
\label{TP}
\end{equation}
Again, for $h=1$ this expression coincides with (\ref{TPSimple}).

\pagebreak

\section{Figures}

\begin{references}
\bibitem{Flatness}  For a review of inflation, see, e.g., A. D. Linde, {\em %
Particle Physics and Inflationary Cosmology} (Harwood Academic, Chur,
Switzerland, 1990); K. A. Olive, Phys. Rep. {\bf 190}, 307 (1990).

\bibitem{OpenInfl}  J. R. Gott, III, Nature (London) {\bf 295}, 304 (1982);
J. R. Gott, III and T. Statler, Phys. Lett. {\bf 136B}, 157 (1984).

\bibitem{Turok1}  M. Bucher, A. S. Goldhaber, and N. Turok, Phys. Rev. D 
{\bf 52}, 3314 (1995); M. Bucher and N. Turok, Phys. Rev. D {\bf 52}, 5538
(1995).

\bibitem{OpenInfl1}  T. Tanaka and M. Sasaki, Phys. Rev. D {\bf 50}, 6444
(1994); K. Yamamoto, T. Tanaka and M. Sasaki, Phys. Rev. D {\bf 51}, 2968
(1995).

\bibitem{LindeO1}  A. D. Linde, Phys. Lett. B {\bf 351}, 99 (1995); A. D.
Linde and A. Mezhlumian, Phys. Rev. D {\bf 52}, 6789 (1995).

\bibitem{Density}  J. Garriga, preprint gr-qc/9602025; J. Garcia-Bellido,
preprint astro-ph/9510029.

\bibitem{Predictions}  A. Vilenkin, Phys. Rev. Lett. {\bf 74}, 846 (1995).

\bibitem{Anthropic}  B. Carter, in I.A.U. Symposium, vol. {\bf 63}, ed. by
M. S. Longair (Reidel, Dordrecht, 1974).

\bibitem{Anthropic1}  B. J. Carr and M. J. Rees, Nature 278, 605 (1979).

\bibitem{Anthropic11}  J. Barrow and F. Tipler, {\em The Antropic
Cosmological Principle} (Clarendon Press, Oxford, 1986).

\bibitem{LB}  A. D. Linde, {\em Particle Physics and Inflationary Cosmology}
(Harwood Academic, Chur, 1990).

\bibitem{Anthropic2}  J. Leslie, {\em Mind} {\bf 101}, 521 (1992).

\bibitem{Fn3}  The thermalization temperature is a convenient reference
point, but of course this choice is arbitrary, and one can use any other
reference temperature.

\bibitem{LLM}  A. D. Linde, D. A. Linde, and A. Mezhlumian, Phys. Rev. D 
{\bf 49}, 1783 (1994).

\bibitem{GBL}  J. Garcia-Bellido, A. D. Linde, and D. A. Linde, Phys. Rev. D 
{\bf 50}, 730 (1994).

\bibitem{MakingPredictions}  A. Vilenkin, Phys. Rev. D {\bf 52}, 3365 (1995).

\bibitem{Uncertainties}  S. Winitzki and A. Vilenkin, Phys. Rev. D {\bf 53},
4298 (1996).

\bibitem{Fn1}  It should be noted that the requirement of
time-reparametrization invariance alone does not fix a unique regularization
procedure. In fact, Linde and Mezhlumian \cite{OtherReg} suggested a
generalization of the $\epsilon $-prescription in which the co-moving volume 
${\cal V}_c$ is replaced by a weighted volume ${\cal V}_q={\cal V}_c\left( 
{\cal V}/{\cal V}_c\right) ^q$, where ${\cal V}$ is the physical volume and $%
q$ is a dimensionless parameter. Although all regularizations belonging to
this one-parameter family are time-reparametrization invariant, it was
argued in \cite{Uncertainties} that the original $\epsilon $-prescription
(which corresponds to $q=0$) has some important advantages: (i) unlike
regularizations with $q>0$, it can be applied to arbitrary inflaton
potentials, and (ii) it gives the ``correct'' answer in some cases where a
certain result is expected on intuitive grounds. We shall, therefore, adopt
the original $\epsilon $-prescription in this paper and only briefly comment
on the results one would obtain from the modified $q\neq 0$ prescriptions in
Sec.\ \ref{CONCL}.

\bibitem{Diffusion}  For a recent review of this stochastic approach to
inflation, see \cite{LLM}.

\bibitem{FnFunction}  The function $f\left( h\right) $ is rather
complicated, 
\[
f\left( h\right) =\left( 1+h\right) ^3\frac{\left( 69-17h^2+6h^3-18h\right)
\left( 1+h\right) ^{\frac 3h}+8\left( h^2-21\right) }{9\left( 9-h^2\right)
\left( 9-4h^2\right) }.
\]
Its explicit form will not be useful for us, and we can use the linear fit (%
\ref{FHFit}) to visualize its behavior.

\bibitem{OtherRegV}  As we noted before \cite{Fn1}, the regularization
procedure of \cite{MakingPredictions} which we use here is not unique, and a
set of alternative prescriptions depending on a parameter $q$ was proposed
in \cite{OtherReg}; the original prescription is obtained for $q=0$.
Analogous calculations can be performed using the alternative procedure. For
very small positive or negative $q$ satisfying $-\left( H_0\tau _{*}\right)
^{-1}\lesssim 3q\lesssim 3-d$, the modified Eq.\ (\ref{VStarRegJ}) is 
\[
{\cal V}_{*}^{(j)}\propto f\left( h^{(j)}\right) \gamma ^{(j)}\left[
a_{*}^{(j)}\right] ^3\exp \left( \frac{3qd}{3-d-3q}H_0\tau _{*}^{(j)}\right)
\epsilon ^{-\frac d{3-d-3q}}.
\]
For larger negative $q$ satisfying $\left| 3q\right| \gtrsim \left( H_0\tau
_{*}\right) ^{-1}$, it becomes 
\[
{\cal V}_{*}^{(j)}\propto f\left( h^{(j)}\right) \gamma ^{(j)}\left( H_0\tau
_{*}^{(j)}\right) ^{-\frac 1q}\epsilon ^{-\frac d{3-d-3q}}.
\]

\bibitem{Adams}  F. C. Adams, Phys. Rev. D {\bf 48}, 2800 (1993).

\bibitem{OtherRegP}  One can compare the probability distribution for $%
\Omega $ obtained using alternative regularization procedures \cite{Fn1}
parametrized by $q$ with that in the $q=0$ case. The relevant formulae for
the thermalized volumes are given above in footnote\ \cite{OtherRegV}. The
allowed range of $q$ is $3q<3-d$, where $d\approx 3$ is given by (\ref
{DSimple}). The behavior of the distribution is similar to the $q=0$ case,
with maxima at $\Omega =0$ and $\Omega =1$ depending on the value of $\mu $,
but the values of $\mu $ separating different regimes become $q$-dependent.
For large negative $q$ that satisfy $\left| 3q\right| \gtrsim \left( H_0\tau
_{*}\right) ^{-1}$, the peak is always at $\Omega =0$, whereas for small $%
\left| q\right| $ in the interval $-\left( H_0\tau _{*}\right) ^{-1}\lesssim
3q\lesssim 3-d$ there are values of $\mu $ for which the peak is at $\Omega
=1$.

\bibitem{FluctO}  The spectrum of density fluctuations is also $\Omega $%
-dependent, but this dependence is negligible on scales small compared to
the curvature radius of the bubble. This range of scales includes the
galactic scale, except for very small values of $\Omega $. Since $\nu _{%
\text{civ}}\left( \Omega \right) $ drops exponentially fast as $\Omega $ is
decreased, we expect that our conclusions will not be substantially modified
by taking into account the $\Omega $-dependence of the fluctuation spectrum.

\bibitem{Fn4}  It should be emphasized that our evaluation of $\nu _{\text{%
civ}}\left( \Omega \right) $ can serve only as a very rough estimate. In
particular, for very low values of $\Omega $, galaxies are formed at a high
redshift ($z\sim z_\Omega $), and their properties may be very different
from those observed in our part of the universe. For example, a higher gas
density in the galaxy can affect the rate of star formation, and thus the
number of inhabitable stellar systems.

\bibitem{Anthropic3}  G. Steigman and J. E. Felten, Space Science Reviews 
{\bf 74}, 245 (1995); A. D. Linde and A. Mezhlumian, in Ref.\ \cite{LindeO1}.

\bibitem{Cosm1}  S. Weinberg, Phys. Rev. Lett. {\bf 59}, 2607 (1987).

\bibitem{Cosm2}  G. Efstathiou, M.N.R.A.S. {\bf 274}, L73 (1995).

\bibitem{Cosm3}  A. Vilenkin, in ``Cosmological Constant and the Evolution
of the Universe'', ed. by K. Sato, T. Suginohara, and N. Sugiyama (Universal
Academy Press, Tokyo, 1996).

\bibitem{BGV}  V. A. Berezin, V. A. Kuzmin, I. I. Tkachev, Phys. Lett. 120B,
91 (1983); Phys. Rev. D 36, 2919 (1987); also in {\em Quantum Gravity}, ed.
by M. A. Markov, V. A. Berezin and V. P. Frolov (World Scientific,
Singapore, 1985), p.781; R. Basu, A. H. Guth, and A. Vilenkin, Phys. Rev. D 
{\bf 44}, 320 (1991).

\bibitem{OtherReg}  A. D. Linde and A. Mezhlumian, Phys. Rev. D {\bf 53},
4267 (1996).
\end{references}
\end{document}